\documentclass[a4paper,12pt]{article}
\usepackage[english]{babel}
\usepackage{jheppub2}
\usepackage{subfig}
\usepackage[T1]{fontenc} 
\usepackage{graphicx}
\usepackage{epsfig}
\usepackage{amsmath}
\usepackage{amssymb}
\usepackage{float}
\usepackage{placeins}
\usepackage{braket}
\usepackage{slashed}
\usepackage{braket}
\usepackage{hyperref}

\allowdisplaybreaks[4]

\title{Renormalization in large-$N$ QCD is incompatible with open/closed string duality}

\author[a]{Marco Bochicchio}

\affiliation[a]{INFN Sezione di Roma, Piazzale Aldo Moro 2, 00185, Rome, Italy}

\emailAdd{marco.bochicchio@roma1.infn.it}

\abstract{  Solving by a canonical string theory, of closed strings for the glueballs and open strings for the mesons, the 't Hooft large-$N$ expansion of QCD is a long-standing problem that resisted all the attempts despite the advent 
of the celebrated gauge/gravity duality in the framework of string theory. We demonstrate that in the canonical string framework such a solution does not actually exist because an inconsistency arises between the renormalization properties of the QCD S matrix at large $N$, recently worked out in \href{https://doi.org/10.1103/PhysRevD.95.054010}{Phys. Rev. D {\bf 95}, 054010}, and the open/closed duality of the would-be string solution. Specifically, the would-be open-string one-loop corrections to the tree glueball amplitudes must be ultraviolet (UV) divergent by the aforementioned renormalization properties, which follow from the QCD asymptotic freedom (AF) and renormalization group. Hence, naively, the inconsistency arises because these amplitudes are dual to tree closed-string diagrams, which are universally believed to be both UV finite -- since they are closed-string tree diagrams -- and infrared (IR) finite because of the glueball mass gap.
In fact, the inconsistency follows from a low-energy theorem of the Novikov-Shifman-Vainshtein-Zakharov (NSVZ) type derived in  \href{https://doi.org/10.1103/PhysRevD.95.054010}{Phys. Rev. D {\bf 95}, 054010} that controls the renormalization in QCD-like theories. The aforementioned inconsistency extends to the would-be canonical string for a vast class of 't Hooft large-$N$ confining asymptotically free QCD-like theories including $\mathcal{N}=1$ SUSY QCD. We also demonstrate that the presently existing SUSY string models with a mass gap based on the gauge/gravity duality-- such as Klebanov-Strassler, Polchinski-Strassler (PS) and certain PS variants -- cannot contradict the above-mentioned results, not even potentially, since they are not asymptotically free. Moreover, we shed light on the way the open/closed string duality may be perturbatively realized in these string models compatibly with a mass gap in the 't Hooft-planar closed-string sector and the aforementioned low-energy theorem because of the lack of AF. Finally, we suggest a noncanonical way-out for asymptotically free QCD-like theories based on topological strings on noncommutative twistor space.    }

\DeclareMathOperator{\Tr}{Tr}
\newcommand{\be}{\begin{equation}}
\newcommand{\ee}{\end{equation}}

\newcommand{\bea}{\begin{eqnarray}}
\newcommand{\eea}{\end{eqnarray}}
\newcommand{\bfig}{\begin{figure}}
\newcommand{\efig}{\end{figure}}
\newcommand{\bc}{\begin{center}}
\newcommand{\ec}{\end{center}}

\begin{document}

\maketitle
\flushbottom

\begin{figure}[t]
\centering
\hspace{0.0cm}\includegraphics[width=.5\textwidth]{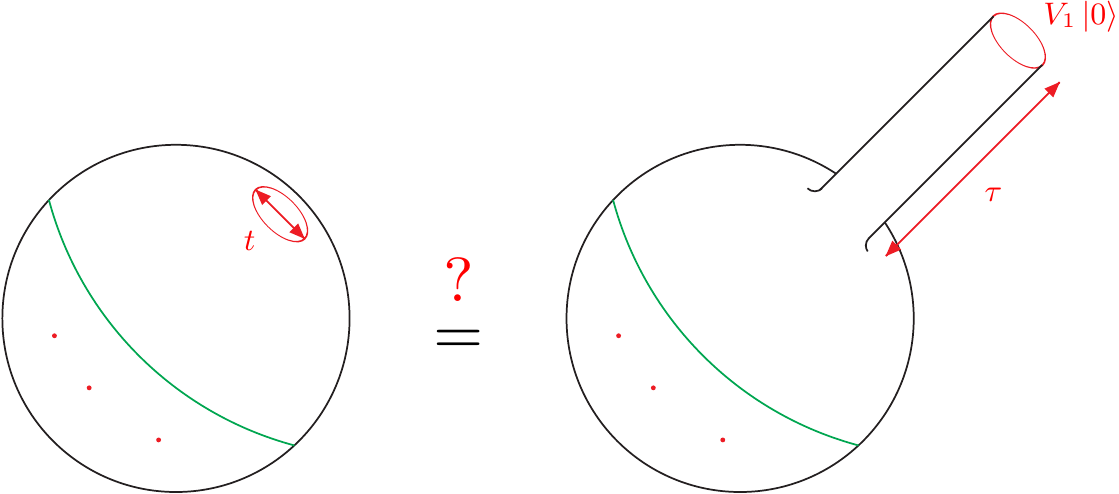}
\hfill
\caption{If a canonical string solution (Subsec. \ref{A}) for large-$N$ QCD exists, the one-loop (hole) open-string diagram in the lhs is a cutoff-independent volume form on the moduli (Subsec. \ref{C}) that, by integrating on the moduli, specifically on $t$, must be UV log divergent  because of the renormalization properties of large-$N$ QCD (Subsec. \ref{B}). By the open/closed duality (Subsec. \ref{C}) the lhs coincides, as a volume form on the moduli, with the tree closed-string diagram in the rhs, where the glueballs propagate in the cylinder of length $\tau=f(t)$, $\tau \rightarrow \infty$ as $t \rightarrow 0$. Thus, the rhs must be divergent, but only by integrating on the moduli, specifically on $\tau$, that is incompatible with the universal belief that is both UV finite -- since it is a tree closed-string diagram (Subsec. \ref{D}) -- and IR finite because of the glueball mass gap (Subsecs. \ref{D} and \ref{E}). In fact, a low-energy theorem (Subsec. \ref{E}) in large-$N$ QCD implies that the rhs is log divergent before integrating on $\tau$ because of the explicit UV log divergence of the operator counterterm, $V_1$, in the boundary state, $V_1 \ket{0}$. Hence, the rhs cannot coincide as a volume form on the moduli with the lhs, and the open/closed duality cannot hold in large-$N$ QCD (Subsec. \ref{E}).}
\label{fig}
\end{figure}

\section{Introduction and conclusions} \label{alpha}

Solving the 't Hooft large-$N$ expansion of QCD \cite{H1} with gauge group $SU(N)$ and $N_f$ quark flavors is a long-standing problem with outstanding implications for the theory of strong interactions and perhaps for the physics beyond the standard model, if any, in case it arises from a new strongly interacting sector. It is universally believed \cite{H1, V0, Mal} that such a solution may be a string theory \cite{VR}, of closed strings for the glueballs and of open strings for the mesons \cite{H1,V0}. \par
Physically, this is the standard picture of quark confinement in which the string world-sheet is identified with a chromo-electric flux tube, mesons are quark-antiquark bound states linked by the chromo-electric flux, and glueballs are closed rings of flux. \par
Specifically, the QCD perturbative expansion in terms of 't Hooft coupling, $g^2=g^2_{YM} N$, can be rearranged in powers of $\frac{1}{N}$ for $N_f$ fixed, in such a way that there is a correspondence at topological level between Riemann surfaces that represent the string world-sheet, with string coupling $g_{Closed} \sim g_{Open}^2 \sim \frac{1}{N}$, and sums of Feynman graphs triangulating the very same topological surfaces \cite{H1, V0}.\par
This topological matching has been suggesting the existence of a canonical string solution \cite{H1,V0,VR,Mal} for large-$N$ QCD and, more generally, for confining asymptotically free QCD-like theories. 
Yet, this string solution has not been found for more than forty years despite the advent of the celebrated gauge/gravity duality \cite{Mal}, which aims to realize it for large-$N$ QCD and QCD-like theories \cite{Mal}. \par
The crucial ingredient of the canonical string ansatz (Subsecs. \ref{A} and \ref{C}), in addition to the topological matching with the Feynman diagrams, is the existence of an auxiliary 2d conformal field theory living on the string world-sheet of fixed topology \cite{VR}, which is employed to define the S-matrix amplitudes \cite{VR} and possibly also the correlators \cite{Mal} in the supposed string solution for 't Hooft large-$N$ QCD. \par
We believe that the aforementioned definition of the canonical string ansatz captures what a string theory -- at least perturbatively in the string coupling -- has been thought to be for the last forty years \cite{Mal, VR}. \par Yet, we may wonder whether string theories in Ramond-Ramond backgrounds \cite{Pol,BVW,B,O} match our definition of the canonical string ansatz \footnote{We would like to thank the referee for raising this point.}. 
There exist two alternatives. \par
If a conformal quantization for Ramond-Ramond backgrounds exists at least in principle, as advocated, for example, in \cite{BVW,B,O} that we agree with, the Ramond-Ramond backgrounds match our definition of the canonical string framework and the arguments in this paper apply to them as well. \par
If a conformal quantization does not exist for Ramond-Ramond backgrounds, the string arguments in this paper do not apply to them, but because, according to our definition, no corresponding canonical string theory exists from the very start as a consequence of the above assumption. \par
The existence of the canonical-string conformal structure has far-reaching consequences -- for example, the equality 
of the $s$- and $t$-channel S-matrix amplitudes \cite{VR} --
and it is an ad hoc assumption that has not been derived by the fundamental principles of large-$N$ QCD as a field theory.\par
In fact, it is not at all obvious that is compatible with the structure of 't Hooft large-$N$ QCD, despite its compatibility has been taken for granted \cite{H1,V0, Mal} for the past forty years. In this paper, for the first time, we investigate the aforementioned compatibility. \par
We demonstrate that such long sought-after string solution does not actually exist in the canonical string framework because an inconsistency arises, starting from the order of $\frac{N_f}{N}$, between the nonperturbative renormalization properties of the large-$N$ QCD S matrix \cite{MBR} -- a consequence of the QCD asymptotic freedom (AF) and renormalization group (RG) \cite{MBR} -- and the open/closed string duality \cite{V} -- a consequence of the world-sheet conformal structure \cite{V} --. \par
The argument is summarized in the caption of Fig. \ref{fig} and described in detail in Subsecs. \ref{D}, \ref{E}. 
It is followed by the technical proof involving a low-energy theorem that controls the renormalization in QCD both perturbatively (Sec. \ref{1}) and nonperturbatively at large $N$ (Sec. \ref{2}). The fundamental premises are recalled in Subsecs. \ref{A}, \ref{B}, \ref{C}. Incidentally, no obstruction to the open/closed duality occurs perturbatively (Secs. \ref{1} and \ref{3}). \par
Similar considerations apply to 't Hooft large-$N$ $\mathcal{N}$=1 SUSY QCD.
In this respect, in the past twenty years several supergravity string models with a mass gap -- such as Klebanov-Strassler (KS), Polchinski-Strassler (PS), and certain PS variants (Sec. \ref{3})-- have been proposed as string duals to supersymmetric gauge theories, on the basis of the gauge/gravity duality. \par
Thus, since these models are believed to be perfectly consistent string theories dual to supersymmetric gauge theories, the question arises whether the two no-go theorems in this paper (Subsec. \ref{E}), if applied to 't Hooft large-$N$ $\mathcal{N}=1$ SUSY QCD, are in fact compatible with their existence. \par
We will show in Sec. \ref{3} that these string models are neither asymptotically free nor their large-$N$ 't Hooft-planar limits are, by computing how they deviate from the universal asymptotics of the operator product expansion (OPE) of $\Tr F^2$ in an asymptotically free gauge theory that enters crucially the nonperturbative version of the low-energy theorem (Sec. \ref{2}): they are either strongly coupled or asymptotically conformal in the ultraviolet (UV).\par
Besides, certain PS variants (Sec. \ref{3}) involve the large-$N$ Veneziano limit \cite{V0} as opposed to the 't Hooft limit. \par 
Moreover, even by limiting ourselves to closed strings only, for which no consistency issue with AF arises in the large-$N$ 't Hooft expansion (Subsec. \ref{D}), no canonical string model presently exists whose two-point correlator for $\Tr F^2$ has been shown to factorize over an infinite tower of massive states and to agree asymptotically in the UV with the result implied by AF in Eq. \ref{11} and \ref{21}, as it must occur in the 't Hooft-planar sector of a confining asymptotically free QCD-like theory (Sec. \ref{3}).\par
Thus, the class of potential string counterexamples of the no-go theorems is presently void. \par
Accordingly, we clarify in Sec. \ref{3}, by means of the perturbative version of the low-energy theorem (Sec. \ref{1}), how the open/closed duality in the aforementioned string models may be in fact compatible at lower perturbative orders, but precisely because of the lack of AF,
with the potential large-$N$ renormalization due to matter fields in the fundamental representation and mass gap in the 't Hooft-planar closed-string sector.\par
Finally, we suggest a noncanonical way-out of the no-go theorems for 't Hooft expansion in asymptotically free QCD-like theories, based on topological strings on noncommutative twistor space (Subsec. \ref{E}).

\section{Fundamental premises and no-go theorems} \label{0}

\subsection{The canonical string framework and the 't Hooft-planar theory} \label{A}

In the canonical string framework for 't Hooft large-$N$ QCD, S-matrix amplitudes are defined in terms of suitable correlators in an auxiliary 2d conformal field theory (Subsec. \ref{C}) living on the string world-sheet with fixed topology, matching the topology of the Feynman diagrams in the 't Hooft expansion \cite{H1,V0,Mal,VR}.\par
The leading contribution at large $N$ is the 't Hooft-planar theory \cite{H1} that consists of planar diagrams with the topology of a punctured sphere in the glueball sector, and of a disk with punctures on the boundary in the meson sector.\par
Since the planar S matrix only involves tree-level interactions \cite{Migdal,Witten}, it must be UV finite.
In planar QCD with massless quarks the UV finiteness is obtained by the renormalization of $g$, in such a way that the planar S matrix is UV finite once it is expressed \cite{MBR} in terms of the planar RG invariant, $\Lambda_{QCD}^P$, that is a nonperturbative quantity, since it vanishes perturbatively. The same UV finiteness must occur in the tree-level string solution, if it exists, which thus must depend on a parameter, $\sqrt {T}$, identified with $\Lambda_{QCD}^P$. For renormalized quark masses, $m^P$, the masses of the pseudo-Goldstone bosons, $M_{PGB}^P$, also occur as parameters in planar QCD. \par
The planar S matrix is not unitary yet, as it consists of tree amplitudes. To restore unitarity \cite{H1,V0} in the S matrix, non-'t Hooft-planar diagrams, which are subleading in $\frac{1}{N}$, must be included. They carry handles and holes (Fig. \ref{fig}), which nonperturbatively involve glueball and meson loops respectively. The question arises whether the nonplanar S matrix is UV finite (Subsec. \ref{B}).

\subsection{Nonplanar renormalization in large-$N$ QCD}  \label{B}

A main result in \cite{MBR} is that nonperturbatively, once the 't Hooft-planar S matrix has been made finite by the planar renormalization common to the YM theory and massless QCD, RG and AF imply that the further nonplanar contributions are UV finite in the YM S matrix, while they are only renormalizable in the S matrix of QCD and QCD-like theories \cite{MBR}. \par
Indeed, in the perturbative YM S matrix, all the nonplanar divergences only affect the scheme-dependent coefficients of the beta function, and thus nonperturbatively can be reabsorbed in a finite renormalization of $\Lambda_{QCD}^P$ \cite{MBR}. \par
Instead, in the perturbative QCD S matrix, non- 't Hooft-planar quark loops from the quark functional determinant are UV divergent, and modify the scheme-
independent first coefficient of the planar beta function. Thus, these divergences cannot be reabsorbed nonperturbatively in a finite renormalization of $\Lambda_{QCD}^P$ \cite{MBR}. \par
Indeed, according to \cite{MBR}, the large-$N$ expansion of the RG invariant in $SU(N)$ QCD, $\Lambda_{QCD}$, in terms of the planar RG invariant, $\Lambda_{QCD}^P$, reads asymptotically up to the order of $\frac{N_f}{N}$:
\bea \label{alpha1}
\Lambda_{QCD} &\sim&  \Lambda \exp\big(-\frac{1}{2\beta^{P}_0 (1+\frac{\beta_0^{NP}}{\beta_0^{P}}) g^2 }\big) \nonumber \\
&\sim&  \Lambda \exp(-\frac{1}{2\beta^{P}_0  g^2 })(1+\frac{\frac{\beta_0^{NP}}{\beta_0^{P}}}{2\beta^{P}_0  g^2 }) \nonumber \\
&\sim& \Lambda^P_{QCD} \big(1+\frac{\beta_0^{NP}}{\beta_0^{P}} \log(\frac{\Lambda}{ \Lambda^P_{QCD}})\big) 
\eea
where $g$ has been eliminated identically in terms of $\Lambda_{QCD}^P$.
The log divergence in the last term of Eq. \ref{alpha1} arises because of AF in the 't Hooft-planar theory, i.e., $\frac{1}{2\beta^{P}_0  g^2 } \sim  \log(\frac{\Lambda}{ \Lambda^P_{QCD}})$,
and the non-'t Hooft-planar change of $\beta_0=\beta^P_0+\beta^{NP}_0$, i.e., $\beta_0^{NP} \neq 0$, due to the matter fields in the fundamental representation, i.e., the quarks.\par
Because of the log divergence, it is impossible to find a renormalization scheme for $g$ that makes $\Lambda_{QCD}$ and $\Lambda_{QCD}^P$ both finite at the same time. \par 
Hence, since the first term in the large-$N$ expansion, i.e., the 't Hooft-planar theory, must be made finite, the renormalization of $\Lambda_{QCD}^P$ must occur necessarily order by order, for keeping $\Lambda_{QCD}$ finite in the $\frac{1}{N}$ expansion \cite{MBR}. \par
As a consequence, the non-'t Hooft-planar diagram in the lhs of Fig. \ref{fig}, which renormalizes at order of $\frac{N_f}{N}$ the $k$-glueball tree amplitude, is UV log divergent \cite{MBR}: perturbatively because of the quark loop (i.e., the hole boundary), and nonperturbatively because of the coupling with the mesons.

\subsection{Open/closed string duality, potential divergences in the open-string sector, and classification of string world-sheet divergences from
the QCD 4d field-theory perspective}  \label{C}

The exact conformal symmetry on the string world-sheet implies the open/closed duality \cite{V}, i.e., the equality for conformally equivalent diagrams of the open- and closed-string partition functions viewed as volume forms on the world-sheet moduli \cite{V}. \par
Specifically, the graph in the lhs of Fig. \ref{fig} is a mixed open/closed string diagram obtained by gluing a disk, $D_k$, with $k$ punctures in the interior, to an annulus, $Annulus(t)$, with modulus $t$. It represents the interaction of closed string states, created at the punctures, with open strings, which end at the appropriate D-brane \cite{V,DV} on the hole boundary. \par
The annulus with modulus $t$ can be mapped by a conformal transformation to a cylinder, $Cylinder(\tau)$, with modulus $\tau$, 
for a function, $\tau= f(t)$, such that $\tau \rightarrow \infty$ as $t \rightarrow 0$ \cite{V,DV}.\par
The resulting graph in the rhs of Fig. \ref{fig} represents a purely closed-string diagram, describing the interaction of the aforementioned closed-string states, 
created at the punctures, with an extra closed string, which propagates in the cylinder, and it is annihilated by the appropriate D-brane closed-string state, $V_1 |0  \rangle$, at the cylinder boundary  \cite{V,DV}.\par
Hence, in a given canonical string theory that is supposed to be exactly conformal invariant on the world-sheet
by definition, the open- and closed-string partition functions, arising respectively in the lhs and rhs of Eq. \ref{local1} from the two graphs in Fig. \ref{fig}, must be equal as volume forms on the world-sheet moduli,
since they live, in fact, on the very same graph, up to the conformal transformation. \par 
This equality is an example of the open/closed string duality, and it holds independently of the technical details of the formulation of the string theory.\par
For example, the open/closed duality applies to string theories defined by nontrivial sigma models \cite{Mal}, to string target spaces that may include extra dimensions and curvature \cite{Mal}, and to any D-brane \cite{V,DV} or string sector \cite{DV} as well, such as the Neveu-Schwarz and Ramond \footnote{See the comment about Ramond-Ramond backgrounds in Sec. \ref{alpha}.} sectors \cite{DV}, provided that the resulting string theory can be quantized consistently with the 2d world-sheet conformal symmetry, which is 
the fundamental assumption in the canonical string framework \cite{Mal,VR}. \par
In fact, the open/closed duality applies even to string theories that may not admit a Lagrangian formulation, provided that they are defined by a 2d conformal field theory.\par
For the diagrams in Fig. \ref{fig} the open/closed duality implies:
\bea \label{local1}
&&\langle D_k(m_i) | Annulus(t) \rangle dm_1\wedge \cdots \wedge dm_i  \wedge \frac{dt}{t} \nonumber \\
&& = \langle D_k(m_i) | Cylinder(\tau) V_1 |0  \rangle   dm_1\wedge \cdots \wedge dm_i  \wedge d\tau 
\eea
for the map, $\tau= f(t)$, \cite{V,DV}. $m_1 \cdots m_i$ are the remaining moduli of the surface obtained gluing the punctured disk to the annulus in the open-string side (lhs), or to the cylinder in the closed-string side (rhs). \par
More specifically, in the canonical string framework, the state, $\langle D_k(m_i) |$, is supposed to be created by a product of closed-string vertex operators \footnote{The original construction of vertex operators in \cite{VR} applied to flat target space, but in principle vertex operators exist in every sector of a canonical string theory, though their actual technical construction may be difficult \cite{BVW,B,O}.}, $\mathcal{V}_{j_i}(z_i,\bar z_i)$, supported at the punctures with world-sheet coordinates
$(z_i,\bar z_i)$:
\bea
\langle D_k(m_i)| 
=\langle 0 | \mathcal{V}_{j_1}(z_1,\bar z_1)  \mathcal{V}_{j_2}(z_2,\bar z_2) \cdots \mathcal{V}_{j_k}(z_k,\bar z_k)| 
\eea
The 2d conformal invariance of the S-matrix amplitudes in the canonical string framework requires that the vertex operators be conformal operators of weight $(1,1)$ \cite{VR,Mal}, whose quantum numbers, compactly labelled by $({j_1}, {j_2} \cdots {j_k})$, include the momenta and quantum numbers of the (on-shell) closed-string states that they create. The integration on the moduli, $m_i$, includes the integration on the coordinates of the vertex operators, $(z_i,\bar z_i)$, possibly after fixing some of the punctures by exploiting the possibly nontrivial automorphism group of the world-sheet. \par
Moreover, in every 2d conformal field theory, the state, $ | Cylinder(\tau) V_1 |0  \rangle$, can be represented
as:
\bea
 | Cylinder(\tau) V_1 |0  \rangle= | \exp(- \tau H_{Closed}) V_1 |0  \rangle  
 \eea
where $H_{Closed}$ is the Hamiltonian on the cylinder, which necessarily exists, since the translations along the $\tau$ direction are included in the conformal group on the cylinder. $ |0  \rangle$ is the closed-string vacuum in the given \cite{DV} closed-string sector. \par
The integration on $\tau$ constructs the tree propagator, $\int^{\infty}_{0} \exp(- \tau H_{Closed}) d\tau = H_{Closed}^{-1}$, of the closed string. The closed-string state on the cylinder boundary, $ V_1 |0  \rangle$, does not \cite{DV}, and may not, depend on $\tau$, because all the $\tau$ dependence must occur via the geometric generator of translations along the cylinder, $H_{Closed}$, in order for $H_{Closed}^{-1}$ to arise from the $\tau$ integration. \par
If the QCD canonical string solution exists, after integrating on the moduli, Eq. \ref{local1} implies the equality between (amputated on-shell) glueball $k$-point 4d correlators in large-$N$ QCD within the leading $\frac{1}{N}$ accuracy:
\bea \label{dual1}
&&\braket{\Tr F^2 \cdots \Tr F^2 }^{1-Open String Loop}_{conn} \nonumber \\
&&=  \braket{ \Tr F^2 \cdots  \Tr F^2 \, V_1}^{Tree Closed String}_{conn} 
\eea
for a certain zero-momentum, possibly nonlocal, scalar gauge-invariant pure-glue operator, $V_1$. \par
Generically, in canonical string theories -- and therefore in the would-be canonical string solution for QCD  -- the potential divergences in the string S matrix may only arise after integrating on the world-sheet moduli that play the role of Schwinger parameters of Feynman diagrams in field theory \cite{W}. \par
Specifically, the integrands, $\langle D_k(m_i) | Annulus (t) \rangle$  and $\langle D_k(m_i) | \exp(- \tau H_{Closed})$ $ V_1 |0  \rangle$, are UV cutoff ($\Lambda)$ independent and pointwise finite on their definition domain, though not necessarily integrable functions of the world-sheet moduli, and only functions of the parameters of the string theory: $T$ and $M_{PGB}^P$ in large-$N$ QCD (Subsec. \ref{A}). \par
Thus, the open-string diagram in the lhs of Eq. \ref{dual1} may only be divergent after integrating on the world-sheet moduli in Eq. \ref{local1}, specifically on $t$. As a consequence, only after integrating on the moduli, specifically on $\tau$, the dual closed-string diagram may be divergent.\par
An important observation is that in the would-be canonical string solution for QCD the potential divergences arising from integrating on the world-sheet moduli must necessarily correspond to potential UV or infrared (IR) divergences in the 4d large-$N$ QCD S matrix or correlators, and vice-versa, since the string world-sheet diagrams are supposed to solve exactly for the large-$N$ QCD S matrix or correlators, whose potential divergences have always a physical 4d interpretation. \par
This does not imply that the string target space is necessarily 4d. In the canonical string framework any string target space consistent with the world-sheet conformal symmetry -- including extra dimensions, target-space curvature, D-branes and so on \cite{V,DV} -- that may potentially solve for QCD is allowed. But we point out that any divergence arising from integrating on the world-sheet moduli in the would-be canonical QCD string must admit a 4d space-time interpretation in QCD viewed as a 4d field theory and vice-versa:
this simple observation is not widely recognized though.\par
Therefore, we classify the potential closed-string and open-string world-sheet divergences in the would-be canonical string solution for QCD by referring to them for short as to UV or IR divergences: this means that UV or IR is the divergence of the 4d large-$N$ QCD S-matrix amplitude or correlator that arises from integrating on the world-sheet moduli of the corresponding string diagram. \par
We reiterate that by no means this implies that the string target space coincides necessarily with the physical 4d space-time, and, indeed, we make no assumption on the nature of the string target space that is completely irrelevant for the arguments in this paper. The same considerations apply to any string theory that is supposed to solve exactly a given 4d field theory at large $N$.

 \subsection{Nonplanar renormalization in large-$N$ QCD versus its would-be canonical string} \label{D}

The UV finiteness of the large-$N$ YM S matrix \cite{MBR} (Subsec. \ref{B}) matches the universal belief that all the consistent closed-string theories are UV finite \cite{W,S}. This matching is remarkable since two very different and largely independent features imply the finiteness: on the gauge side, AF and RG  \cite{MBR}, on the string side, conformal symmetry on the string world-sheet and, as it is universally believed \cite{W,S}, modular invariance underlying closed-string diagrams. \par
However, in the would-be large-$N$ QCD string solution, there is a tension between the UV log divergence of the $k$-glueball amplitude coupled to mesons in the lhs of Eq. \ref{dual1} (Subsec. \ref{B}) and the universally believed UV finiteness \cite{W,S} of the geometrically planar \cite{V} diagram in the rhs of Eq. \ref{dual1} that, being both a tree and a closed-string diagram, should be UV finite. \par
There exist solvable string examples \cite{DV} that bear deep analogies with the issue considered here, where the canonical formulation of string theory resolves the aforementioned tension \cite{V}: as recalled in Sec \ref{1}, in \cite{DV} the UV divergence of the open-string diagram in the lhs of  Eq. \ref{dual1} is mapped by the open/closed duality into an IR divergence of the dual tree closed-string diagram in the rhs of Eq. \ref{dual1}, according to the relation $\tau \rightarrow \infty$ as $t \rightarrow 0$ in Eq. \ref{local1} \cite{V}. \par
Yet, the IR divergence may only occur if the tree closed-string theory has no mass gap. Therefore, this cannot work for large-$N$ QCD \cite{MBng}
if it is assumed, on the basis of the overwhelming numerical evidence \cite{Me1,Me2,L1,L2}, that large-$N$ QCD has a mass gap in the 't Hooft-planar glueball sector. 

\subsection{The no-go theorems}  \label{E}

Therefore, assuming that the closed-string solution for large-$N$ QCD provides an UV finite tree diagram in the rhs of Eq. \ref{dual1}, as it is universally believed \cite{W,S} and occurs in all the presently-known consistent string models for which there is an explicit realization of the open/closed duality \cite{DV}, we get immediately the first no-go theorem \cite{MBng}:
the open/closed duality in a would-be canonical string solution for the 't Hooft large-$N$ QCD S matrix is incompatible with the UV finiteness of the closed-string trees, RG, AF and the mass gap in the planar glueball sector jointly. \par
The proof of the first no-go theorem is essentially a tautology, given the renormalization properties of large-$N$ QCD: on the closed-string side of the duality there is neither an IR divergence 
because of the glueball mass gap, nor an UV divergence because of the assumption on the UV finiteness of the closed-string tree, while 
on the open-string side there is an UV divergence due to the non-'t Hooft-planar renormalization -- a consequence of AF and RG in large-$N$ QCD (Subsec. \ref{B}) --. This contradicts the open/closed duality, since a closed-string diagram that is both UV and IR finite cannot be dual to an UV divergent open-string diagram (Subsec. \ref{C}).\par
However, by assuming the mass gap, the only logically possible alternative is that the rhs of Eq. \ref{dual1} may be UV log divergent rather than IR log divergent. Accordingly, the potential inconsistency would be resolved by contradicting the universally believed UV finiteness of the tree closed-string diagrams. \par
We find out what actually happens in QCD -- and in asymptotically free QCD-like theories with a mass gap in the 't Hooft-planar glueball sector -- by computing explicitly the operator $V_1$ in Eq. \ref{dual1},
both in perturbation theory and nonperturbatively in the large-$N$ 't Hooft expansion, by means of a low-energy theorem of the Novikov-Shifman-Vainshtein-Zakharov (NSVZ) type derived in \cite{MBR}. \par
In a perturbatively massless QCD-like theory \cite{MBR} (Sec. \ref{1}), the open/closed duality is consistent with perturbation theory at order of $g_{YM}^2$. Indeed, the insertion in Eq. \ref{dual1} of the operator provided by Eq. \ref{Th}, $V_1= \frac{1}{2} \int \Tr F^2 d^4x$, is both UV and IR log divergent because of the perturbative conformal symmetry at order of $g_{YM}^2$ (Eq. \ref{div}). Besides, $V_1$ is $\Lambda$ independent. This is consistent with the $\Lambda$ independence and pointwise finiteness (Subsec. \ref{C}) of the integrand in the rhs of Eq. \ref{local1}.\par
Similarly, in string models that in the 't Hooft-planar closed-string sector are asymptotically conformal in the UV and have a mass gap, the open/closed duality may be realized
at lower orders consistently with the perturbative version of the low-energy theorem (Sec. \ref{3}). \par
Instead, at order of $\frac{N_f}{N}$ in large-$N$ QCD with massless or massive quarks (Sec. \ref{2}), the operator provided by Eqs. \ref{dual1} and \ref{low}, $V_1=  N \beta_0^{NP}  \log(\frac{\Lambda}{ \Lambda^P_{QCD}}) \int \Tr F^2 d^4x+ $non-local  UV-finite terms, contains a counterterm which diverges as $\Lambda \rightarrow \infty$ before its insertion in the correlator in Eq. \ref{dual1}, while the insertion of $ \int \Tr F^2 d^4x$ is, in fact, UV finite because of AF (Eq. \ref{OPE1}). This is incompatible with the $\Lambda$ independence and pointwise finiteness (Subsec. \ref{C}) of the string integrands in both sides of the supposed duality in Eq. \ref{local1}. \par
A second stronger no-go theorem, based only on UV arguments, follows: RG and AF are incompatible with the open/closed duality in a would-be canonical string solution -- that thus does not exist -- for the 't Hooft large-$N$ QCD S matrix. The incompatibility extends to a vast class \cite{MBR} of asymptotically free 't Hooft large-$N$ QCD-like theories including $\mathcal{N}=1$ SUSY QCD.\par
\emph{A posteriori}, in certain topological string theories on noncommutative twistor space, which we proposed \cite{MBH} as candidates for the large-$N$ QCD S matrix, the inconsistency is naturally avoided: the one-loop effective action gets noncanonical contributions from string instantons wrapping geometrically planar surfaces with any number of holes, in such a way that the mechanism which leads to the inconsistency is spoiled (see Sec. 4 in \cite{HEP2017}).

\section{Consistency of open/closed duality with perturbative renormalization in QCD-like theories} \label{1}

Initially, we suppose that the open-string solution reproduces perturbatively a QCD-like theory, i.e., the string coupling is identified with the gauge coupling, $g_{Open} \sim g_{YM}$ \cite{DV}. We verify that the low-energy theorem \cite{MBR} at order of $g_{YM}^2$ in a perturbatively massless QCD-like theory, and thus in QCD with massless quarks, is compatible with the open/closed duality. \par
Firstly, we write the low-energy theorem for a two-point correlator with the canonical normalization of the action:
\bea \label{pert}
 \frac{\partial\braket{ F^2(z) F^2(0)}}{\partial \log g_{YM}} &=& \int \braket{ F^2(z) F^2(0)  \Tr F^2(x)} \nonumber \\
 & - & \braket{  F^2(z) F^2(0)} \braket{\Tr F^2(x)}d^4x
\eea
Secondly, we employ a result in \cite{K,Ch} for the OPE of the operator $F^2(x)\equiv 2 \Tr F^2(x)$ -- that is well defined at tree level -- at order of $g_{YM}^2$:
\bea \label{OPE}
F^2(z) F^2(0) &\sim& \frac{N^2-1}{z^8}  \frac{48}{\pi^4} \nonumber \\
&& \big(1- 4 \beta_0 g_{YM}^2 (\log\frac{1}{|z| \mu}- \log(\frac{\Lambda}{\mu})) 
+ \cdots \big) \nonumber \\
&+& \frac{1}{z^4} \frac{4 \beta_0}{\pi^2} g_{YM}^2 F^2(0)
\eea
where we omitted the finite parts in the dots. Substituting Eq. \ref{OPE} in the lhs of Eq. \ref{pert}, we get:
\bea \label{Th}
&&2 \big[ \braket{ F^2(z) F^2(0)} \big]^{1- Loop} \nonumber \\
&&= \frac{1}{2}  \int \big[ \braket{  F^2(z) F^2(0)  F^2(x)} \big]^{Order \, of \, g_{YM}^2 }d^4x
\eea
We evaluate the divergent parts in Eq. \ref{Th} employing Eq. \ref{OPE} and the OPE in the rhs:
\bea \label{div}
&& 2 \big[ \frac{N^2-1}{z^8}  \frac{48}{\pi^4} 4\beta_0 g_{YM}^2 (\log(|z| \mu) + \log(\frac{\Lambda}{\mu})+\cdots) \big]_{div} \nonumber \\
&&= 2  \frac{1}{2} \int \braket{  F^2(z) \frac{1}{x^4} \frac{4 \beta_0}{\pi^2} g_{YM}^2 F^2(0)}^{Tree}d^4x \nonumber \\
&& =  \braket{  F^2(z) F^2(0) }^{Tree}  \int \frac{4 \beta_0}{\pi^2} g_{YM}^2  \frac{1}{x^4} d^4x \nonumber \\
&& =  \frac{N^2-1}{z^8}  \frac{48}{\pi^4}  8 \beta_0 g_{YM}^2  \log(\frac{\Lambda}{\mu})
\eea
where the factor of $2$ in the rhs occurs because $x$ may be close to $0$ or to $z$.\par
Comparing with Eq. \ref{dual1}, we interpret Eqs. \ref{Th} and \ref{div} in terms of open/closed duality. Eq. \ref{Th} constructs explicitly the closed-string operator, $V_1= \frac{1}{2} \int \Tr F^2 d^4x$ in the rhs, which enters the closed-string side of the duality.\par
The lhs, that represents the open-string side of the duality, is log divergent at one loop in perturbation theory because of the anomalous dimension, $\gamma_0=2 \beta_0$, of $ \Tr F^2$. The divergence is both UV and IR because of the conformal symmetry of perturbation theory at order of $g^2_{YM}$.
This is matched by the UV log divergence of a one-loop open-string diagram in the
string examples \cite{DV} due to the same nontrivial one-loop beta-function coefficient, $\beta_0$. \par
In the rhs of Eq. \ref{Th},
the insertion of $\Tr F^2(x)$ at zero momentum is both UV and IR divergent, as predicted by open/closed duality, because of the conformal symmetry of the OPE at the lowest nontrivial order. In agreement with the IR divergence in the rhs, in all the consistent string models in \cite{DV} there is an IR log divergence in the closed-string sector (i.e., in the gravity sector)
due to a massless dilaton \cite{DV}, which is the string field dual to $\Tr F^2(x)$, that reproduces the very same beta function on the gravity side. \par
Most importantly, the log divergence arises in the rhs of Eq. \ref{Th} only after inserting the well-defined (at tree level) $\Lambda$-independent operator, $\int \Tr F^2 d^4x$, in the v.e.v.. This is consistent with the $\Lambda$ independence of the closed-string integrand in Eq. \ref{local1}, as the log divergence may arise in the string diagram only by integrating on the moduli, specifically on $\tau$ (Subsec. \ref{C}).\par
Finally, everything that we mentioned is compatible with the first no-go theorem (Subsec. \ref{E}), as there is no mass gap both in perturbation theory and in the aforementioned string models. \par
The perturbative computation also suggests that non-asymptotically-free theories (see Sec. 6 in \cite{MBR}), which are asymptotically conformal in the UV or in the IR, may be compatible with the duality. Specifically, there may be no perturbative obstruction to the open/closed duality in string models that, in their 't Hooft-planar closed-string sector, are asymptotically conformal in the UV and have a mass gap (Sec. \ref{3}).

\section{Inconsistency of open/closed duality with renormalization in large-$N$ QCD-like theories} \label{2}

The situation is drastically different if it is assumed that the string solution reproduces nonperturbatively 't Hooft large-$N$ expansion, i.e., $g_{Closed} \sim \frac{1}{N}$. In this case, in a perturbatively massless QCD-like theory, the low-energy theorem, within the leading $\frac{1}{N}$ accuracy, reads \cite{MBR}:
\bea \label{low}
&&\big[\braket{\Tr F^2 \cdots \Tr F^2 }^{NP}\big]_{div}  \nonumber \\
&&=\frac{N\beta^{P}(g) \Lambda_{QCD}^{NP}}{g^3\Lambda_{QCD}^{P}} \int
\braket{ \Tr F^2 \cdots  \Tr F^2}^{P} \braket{\Tr F^2(x)}^{P} \nonumber \\
&&-\braket{ \Tr F^2 \cdots  \Tr F^2  \Tr F^2(x)}^{P} d^4x \nonumber \\
&& \equiv \braket{ \Tr F^2 \cdots  \Tr F^2 [V_1^{P}]_{div}}^{P}  - \braket{ \Tr F^2 \cdots  \Tr F^2}^{P} \braket{[V_1^{P}]_{div}}^{P}
\eea
Eq. \ref{low} has precisely the structure to match Eq. \ref{dual1}, and it computes its divergent parts in both sides \cite{MBR} due to the renormalization of $\Lambda_{QCD}^P$, with:
\bea
[V_1^{P}]_{div}&=& 
-\frac{N \Lambda_{QCD}^{NP}}{\Lambda_{QCD}^{P}} \frac{\beta^{P}(g)}{g^3} \int {\Tr F^2} d^4x \nonumber \\
&=& N [\beta_0^{NP}  \log(\frac{\Lambda}{ \Lambda^P_{QCD}})+\cdots]  \int \Tr F^2 d^4x
\eea
Moreover, in large-$N$ QCD, $V_1$ can be derived directly from its definition in Eq. \ref{dual1} by expanding at the order of $\frac{N_f}{N}$ the QCD action after integrating  the quark fields \footnote{We employ the Euclidean notation. The analytic continuation to Minkowski space-time is understood.}, $V_1= N_f \Tr \log \frac{\slashed D(A)+Z^P_m m^P}{ \slashed D(0)+Z_m^P m^P} = N \beta_0^{NP}  \log(\frac{\Lambda}{ \Lambda^P_{QCD}})  \int \Tr F^2 d^4x \ +$ nonlocal  UV finite terms,
with $Z_m^P$ defined by the expansion, $Z_m \sim [\log(\frac{\Lambda}{ \Lambda_{QCD}})]^{- \frac{\gamma_{0m}}{2 \beta_0}}$ $ \sim Z_{m}^P (1 + \cdots)$, with $\beta_0=\beta_0^P+\beta_0^{NP}$ and $\gamma_{0m}= \frac{3}{(4 \pi)^2} (1-\frac{1}{N^2})$, where the dots both in $Z_m$ and in $[V_1^{P}]_{div}$ stand for subleading UV log-log divergences \cite{MBR}. \par
Remarkably, within the leading-log accuracy, $ - [V_1^{P}]_{div}$ $ = - [V_1]_{div}$ is the local counterterm, due to the coupling with the mesons \cite{MBR} or to the quark loops from the quark functional determinant, that produces the non-'t Hooft-planar correction to $\beta^P_0$ \cite{MBR}. 
Besides, we will see below that the insertion of $\int {\Tr F^2} d^4x$ in the rhs of Eq. \ref{dual1} is UV finite.
Thus, in QCD both with massless and massive quarks, the UV finite nonlocal terms in $V_1$ may contribute in the rhs of Eq. \ref{dual1} at most UV log-log divergences. \par
The obstruction to the open/closed duality is the UV log divergence of the local part in $V_1$ before its insertion in the rhs of Eq. \ref{dual1}, as opposed to the log-log divergences that occur after the insertion. \par
Indeed, by transferring into the would-be string solution the QCD result for $V_1$ in the rhs of Eq. \ref{local1}, 
$\langle D_k(m_i) | $ $ \exp(- \tau H_{Closed}) V_1 |0  \rangle$ is $\Lambda$ dependent and, as a function of the moduli, pointwise UV log divergent as $\Lambda \rightarrow \infty$, since $V_1|0\rangle$ is created from the closed-string vacuum, $|0\rangle$, by an operator, $V_1$, that contains a log-divergent counterterm. This contradicts the $\Lambda$ independence and pointwise finiteness of the integrand in the rhs side of Eq. \ref{local1}, and thus in both sides (Subsec. \ref{C}). A similar argument holds for $\mathcal{N}=1$ SUSY QCD \cite{MBR}. \par
Naively, there is a simple explanation for this obstruction to the open/closed duality: since the closed-string tree diagrams are UV finite, the only way for the rhs of Eq. \ref{dual1} to be UV log divergent 
is that the boundary state, $V_1 \ket 0$, is explicitly UV log divergent, i.e., it is not well defined in the closed-string theory. \par
Indeed, according to the naive expectation, the $\frac{\beta^{P}(g)}{g^3}$ $  \int {\Tr F^2} d^4x$ insertion is UV finite, as it follows from the asymptotically free RG-improved OPE worked out in \cite{MBN,MBM} within the leading and next-to-leading log accuracy, and afterwards in \cite{R} within the leading-log accuracy:
\bea \label{OPE1}
\beta_0 F^2(z) \beta_0 F^2(0) &\sim& (1-\frac{1}{N^2}) \frac{1}{z^8}  \frac{48\beta_0^2}{\pi^4}   (\frac{1}{\beta_0\log(\frac{1}{z^2 \Lambda^2_{QCD}})})^{2} \nonumber \\
&+& \frac{1}{z^4} \frac{4 \beta^2_0}{\pi^2}  (\frac{1}{\beta_0\log(\frac{1}{z^2 \Lambda^2_{QCD}})})^{2} \frac{\beta_0}{N} F^2(0)
\eea
as opposed to lowest-order perturbation theory that is asymptotically conformal in the UV.\par
Specifically, the UV divergent integral in perturbation theory in the rhs of Eq. \ref{div}, $\int  \frac{1}{x^4} d^4x$, becomes the UV convergent integral in the 't Hooft-planar theory, $\int  \frac{1}{x^4} $ $ \frac{1}{\log^2(\frac{1}{x^2 \Lambda^2_{QCD}})} d^4x$ in the rhs of Eq. \ref{low}, that follows from Eq. \ref{OPE1}.\par
By summarizing, if the canonical string solution exists, the lhs of Eq. \ref{local1} is $\Lambda$ independent, since it is a function only of $T$ and $M_{PGB}^P$, but after integrating on the moduli, specifically on $t$, is UV log divergent because of the large-$N$ QCD renormalization properties. \par
On the contrary, by the large-$N$ QCD computation based on Eqs. \ref{dual1} and \ref{low}, the rhs of Eq. \ref{local1} is $\Lambda$ dependent and UV log divergent as $\Lambda \rightarrow \infty$ before integrating on the moduli, specifically on $\tau$. Thus, Eq. \ref{local1} cannot hold, and the canonical string solution for 't Hooft large-$N$ QCD does not exist. 

\section{Open/closed duality in presently existing string models with a mass gap} \label{3}

\subsection{Fundamental assumptions of the no-go theorems versus KS and PS string models} \label{a}

As it is clear from the proof of both the no-go theorems (Subsec. \ref{E} and Sec. \ref{2}), the consistency issue with the open/closed duality arises because of the UV features of  't Hooft large-$N$ QCD that we summarize as follows. \par
't Hooft $SU(N)$ QCD is asymptotically free (property 1) and so it is its 't Hooft-planar limit as well (property 2).
In $SU(N)$ QCD, the coupling with the matter fields in the fundamental representation, i.e., the quarks, changes the first coefficient of the beta function (property 3) with respect to its 't Hooft-planar limit, and affects the non-'t Hooft-planar corrections to the S matrix.\par
The properties 1 and 2 imply that the UV of both theories is under control by means of the nonperturbative RG resummation of perturbation theory (Subsec. \ref{A}). Specifically,
both $\Lambda_{QCD}$ and $\Lambda_{QCD}^P$ are well defined, and $\Lambda_{QCD}$ can be expanded in terms of $\Lambda_{QCD}^P$ (Subsec. \ref{B}). \par
Moreover, as a result of the property 3, the aforementioned expansion is log divergent starting from the order of $\frac{N_f}{N}$ (Subsec. \ref{B}) because of the property 2. 
In turn, this implies that the second version of the low-energy theorem (Sec. \ref{2}) applies.\par
Last but not least, both QCD and its 't Hooft-planar limit are believed to exist as UV complete 4d gauge theories, again because of AF. \par
The same considerations apply verbatim to 't Hooft large-$N$ $\mathcal{N}=1$ SUSY QCD in the confining/Higgs phase (see Sec. 6 in \cite{MBR}). \par
In this respect, in the past twenty years several supergravity string models with a mass gap, such as KS \cite{KS} and PS \cite{PS}, have been proposed as string duals to supersymmetric gauge theories, not only for the S matrix but for the correlators as well \cite{Mal}, on the basis of the celebrated gauge/gravity duality \cite{Mal}. \par
Thus, the question arises whether their existence is in fact compatible with the no-go theorems in this paper, if applied to 't Hooft large-$N$ $\mathcal{N}=1$ SUSY QCD.\par As we will show momentarily, both KS and PS do not satisfy the properties 1 and 2, i.e., neither they are asymptotically free nor their 't Hooft-planar limits are. Hence, their existence cannot contradict the two no-go theorems, not even potentially.\par
Specifically, KS is strongly coupled in the UV (Subsec. \ref{b}), while PS is asymptotically conformal in the UV \cite{PS} (Subsec. \ref{c}). \par
Thus, these models may be string duals to $\mathcal{N}=1$ SUSY YM theory, as usually referred to in the literature, only in the IR, an important specification -- known to experts \cite{S1,PS}, but not widely recognized -- that plays a key role for the arguments in this paper. \par
Nevertheless, it is interesting to explore in these string models and in certain PS variants \cite{PS} the interplay between potential large-$N$ renormalization due to matter fields in the fundamental representation, mass gap in the 't Hooft-planar closed-string sector, and the open/closed string duality (Subsec. \ref{d}).\par

\subsection{KS is not asymptotically free} \label{b}

Firstly, we examine KS more closely. The first hint that KS is not asymptotically free comes from the original KS paper \cite{KS}: since the gauge group in KS is $SU(N+M) \times SU(N)$, KS involves two gauge couplings that are independent a priori but, in fact, related a posteriori (see Subsec. 2.4 of \cite{KS}). For only one of the gauge couplings occurring in KS, an asymptotically free beta function is obtained. The other gauge coupling flows to strong coupling in the UV generating a duality cascade (see Sec. 3 of \cite{KS}). \par 
The combined effect of the two RG flows in the UV can be evaluated by examining the asymptotic behavior of the two-point correlator
of $\Tr F^2$ in KS \footnote{In the string (supergravity) description, the $\Tr F^2$ correlator is computed \cite{K1,K2} by taking functional derivatives of the effective action with respect to its dual field, the dilaton.} compared with the universal asymptotics of the $\Tr F^2$ correlator in an asymptotically free gauge theory. \par
This comparison has been performed in \cite{MBM,MBN} leading to the following results. In an asymptotically free gauge theory, within the leading-log accuracy:
\begin{equation} \label{11}
\langle \Tr F^2(x) \Tr F^2(0) \rangle \sim \frac{g^4(x)}{x^8} \sim \frac{1}{x^8 \log^2(x^2 \Lambda_{QCD}^2)}
\end{equation}
that in momentum space, up to contact terms, reads:
\begin{equation} \label{21}
\langle \Tr F^2(p) \Tr F^2(-p) \rangle \sim p^4 g^2(p) \sim \frac{p^4}{\log(\frac{ p^2}{\Lambda_{QCD}^2})}
\end{equation}
where $g^2(.)$ is the running coupling, as worked out in  \cite{MBM,MBN} and later confirmed in \cite{R,MBR}.
Indeed, the UV asymptotics in Eqs. \ref{11} and \ref{21} is fixed universally by the anomalous-dimension first coefficient of $\Tr F^2$, $\gamma_0=2 \beta_0$, that is determined by the first coefficient of the beta function, $\beta_0$. \par
Instead, in KS \cite{K1,K2}:
\begin{equation} \label{KS0}
\langle \Tr F^2(x) \Tr F^2(0) \rangle  \sim \frac{\log^2(x^2 \mu^2)}{x^8 }
\end{equation}
that in momentum space, up to contact terms, reads \cite{K1,K2}:
\begin{equation} \label{KS}
\langle \Tr F^2(p) \Tr F^2(-p) \rangle \sim p^4 \log^3(\frac{p^2}{\mu^2})
\end{equation}
as discussed in \cite{MBM,MBN}. The cubic-log growth in the UV in the correlator of $\Tr F^2$ does not correspond to any local 4d gauge theory that we are aware of.  
In fact, there are doubts that such a theory -- strongly coupled in the UV -- can actually exist as an UV complete 4d local gauge theory independently of its string realization \footnote{The puzzling asymptotic behavior in Eqs. \ref{KS0} and \ref{KS} was reported in \cite{K1,K2}. Initially, in \cite{K1} an interpretation was suggested in order to reconcile KS with a local 4d gauge theory by proposing that, after rescaling by the variable number of degrees of freedom along the duality cascade \cite{K1}, $N_{eff} \sim \log(x^2 \mu^2)$, Eq. \ref{00}
would hold in KS after the rescaling. Accordingly, KS would be asymptotically conformal in the UV after the rescaling. If this were the case, our considerations about PS in Subsec. \ref{c} would apply to KS as well. But there is a strong argument against this interpretation on the basis of further findings in \cite{K2}. Indeed, it was found in \cite{K2} that the asymptotic behavior of the trace part of the energy-momentum tensor is:
\begin{equation} \label{KS00}
\langle  T^{\alpha}_{\alpha}(x) T^{\alpha}_{\alpha}(0) \rangle  \sim \frac{\log(x^2 \mu^2)}{x^8 }
\end{equation}
which differs from Eq. \ref{KS0} by one power of the logarithm, contrary to the $\mathcal{N}=1$ local gauge theory, where they should coincide because the trace part of the correlator is proportional to the conformal anomaly that is proportional again to $\Tr F^2$ up to terms that are, at most, less relevant in the UV. Hence, independently of the aforementioned rescaling of the degrees of freedom, Eqs. \ref{KS0} and \ref{KS00} are inconsistent with the $\mathcal{N}=1$ local gauge theory in the ultraviolet. Even if we ignore this issue, the following further inconsistency arises with the interpretation that KS is conformal in the UV after the rescaling. Indeed, even after the rescaling, the trace anomaly does not vanish, in such a way that the theory cannot be conformal in the UV, but the asymptotic behavior for the two-point correlator of the trace anomaly becomes:
\begin{equation} \label{KS000}
\langle  T^{\alpha}_{\alpha}(x) T^{\alpha}_{\alpha}(0) \rangle  \sim \frac{1}{\log(x^2 \mu^2) x^8 }
\end{equation}
which disagrees anyway with the result implied by AF in Eq. \ref{11}. In either case, by comparing the conformal anomaly correlator in Eqs. \ref{KS00} and \ref{KS000} with Eq. \ref{11}, KS is conclusively neither an asymptotically free nor an asymptotically conformal local gauge theory.}.\par Specifically, the KS asymptotics in Eq. \ref{KS0} disagrees with the nonperturbative asymptotics in Eq. \ref{OPE1} that implies the UV finiteness of the operator insertion of $\frac{\beta^{P}(g)}{g^3}$  $\int \Tr F^2(x) d^4x$, which enters the nonperturbative low-energy theorem in Sec. \ref{2}. Moreover, the KS asymptotics in Eq. \ref{KS0} also disagrees with the perturbative asymptotics in Eq. \ref{OPE} that enters the perturbative low-energy theorem in Sec. \ref{1}. \par
Therefore, no version of the low-energy theorem in Secs. \ref{1} and \ref{2} applies to KS, and the existence of KS is not relevant for the implications of this paper. 

\subsection{PS is not asymptotically free} \label{c}

Secondly, we examine PS more closely, which, as opposed to KS, has a clear 4d gauge-theory interpretation.
On the field-theory side, PS is $\mathcal{N}=4$ SUSY YM theory but with its three $\mathcal{N}=1$ chiral supermultiplets in the adjoint representation with masses on the order of $M$, for breaking the supersymmetry to $\mathcal{N}=1$ \cite{PS}. For large $M$, it is argued to reproduce in the IR $\mathcal{N}=1$ SUSY YM theory \cite{PS} \footnote{'t Hooft-planar $\mathcal{N}=1$ SUSY YM theory is believed to have infinitely raising Regge trajectories. It is unclear \cite{PS} which the actual PS spectrum is.}. As a consequence, the crucial difference with $\mathcal{N}=1$ SUSY YM theory is in the UV: because the beta function vanishes identically, the gauge coupling does not run, and the theory is UV finite, as no divergent mass renormalization occurs because of the soft breaking of the $\mathcal{N}=4$ SUSY \cite{PS}. Hence, PS is asymptotically conformal in the UV \cite{PS}.  \par
Thus, the UV asymptotics of the $\Tr F^2$ correlator in PS \footnote{By a standard argument reported in Subsec. 2.3 of \cite{MBM}, the anomalous dimension of
$\Tr F^2$ is $\gamma(g)= g \frac{\partial}{\partial g}(\frac{\beta(g)}{g})$. Thus, it vanishes if the beta function vanishes identically.} is \cite{MBM,MBN}:
\begin{equation} \label{00}
\langle \Tr F^2(x) \Tr F^2(0) \rangle  \sim \frac{1}{x^8 }
\end{equation}
that in momentum space, up to contact terms, reads \cite{MBM,MBN}:
\begin{equation} \label{000}
\langle \Tr F^2(p) \Tr F^2(-p) \rangle \sim p^4 \log(\frac{p^2}{\mu^2})
\end{equation}
that differs by a factor of the square of a log from the asymptotically free case in Eqs. \ref{11} and \ref{21}. Analogously, in PS as compared to an asymptotically free gauge theory, a mismatch by fractional powers of a log affects in general, for example, the UV OPE of twist-two operators \cite{MBN,MBR}, which dominate the deep inelastic scattering. \par
As a consequence, S-matrix amplitudes in PS have been argued \cite{PS1} to scale as powers of the energy, as expected in an asymptotically free theory, up to the aforementioned mismatch by fractional powers of logarithms. \par

\subsection{Perturbative open/closed duality in PS and certain PS variants from the low-energy theorem} \label{d}

PS is a natural string model that may include large-$N$ $\mathcal{N}=1$ SUSY YM theory in the IR, since its existence as a string theory is a natural extension of the string version of large-$N$ $\mathcal{N}=4$ SUSY YM theory, which is universally believed to exist on the basis of the Maldacena conjecture \cite{Mal}.\par
As a canonical string theory, PS is an UV finite theory of closed strings only, and so it is its 't Hooft-planar limit as well. Incidentally, for closed strings only, no consistency problem occurs anyway (Subec. \ref{D}), even in the asymptotically free case of $\mathcal{N}=1$ SUSY YM theory, whose large-$N$ 't Hooft S matrix is, according to \cite{MBR}, UV finite. \par
Thus, PS is an interesting starting point to investigate the open/closed string duality for non-asymptotically-free theories.\par
In this respect, we may couple PS to a fixed number of SUSY matter fields in the fundamental representation with the aim to get $\mathcal{N}=1$ SUSY QCD in the IR. If this gauge-theory model exists as a canonical string theory in the large-$N$ 't Hooft expansion, it is a theory of open and closed strings, and thus we can investigate the open/closed duality in this new setting.\par
Both PS and its 't Hooft-planar limit with the SUSY matter fields in the fundamental representation depend on fixed parameters, $g$, and the masses, $M$, of the $\mathcal{N}=1$ chiral supermultiplets in the adjoint representation, which do not get UV divergent renormalizations, since the 't Hooft-planar beta function is not affected by the presence of the matter fields in the fundamental representation and vanishes as well. \par
This also has deep consequences for the structure of the OPE in the 't Hooft-planar sector of PS in the UV.\par
Indeed, in 't Hooft-planar PS, Eq. \ref{00} holds asymptotically in the UV, which coincides with Eq. \ref{OPE} asymptotically in the UV with $\beta_0=\beta^P_0=0$.
As $g$ is now a fixed parameter that we can choose as small as we like, the first version of the low-energy theorem (Sec. \ref{1}) applies to PS by means of a double expansion in $g^2$ and $\frac{1}{N}$. \par
Firstly, the low-energy theorem can be employed to verify perturbatively in $g^2$ the UV finiteness of PS, and of the 't Hooft-planar limit of PS plus SUSY matter fields as well, because no UV divergence occurs in both sides of Eq. \ref{Th}, since the beta function vanishes.
Besides, the rhs in Eq. \ref{Th} is IR finite because of the assumed large-$N$ $\mathcal{N}=1$ SUSY YM mass gap in the IR.\par
Secondly, the low-energy theorem can be employed to verify, perturbatively at lower orders in $g^2$, the open/closed duality in PS coupled to matter fields in the fundamental representation.\par
Indeed, the non-'t Hooft-planar beta function is different from zero, $\beta_0^{NP} \neq 0$, because of the matter fields in the fundamental representation.\par
Yet, the open/closed duality is perturbatively compatible at lower orders with the first version of the low-energy theorem in Eqs. \ref{OPE}, \ref{Th}, \ref{div}, with $\beta_0=\beta_0^{NP} \neq 0$, because of the UV divergence in the rhs of Eqs. \ref{Th} and \ref{div} induced by the perturbative asymptotically conformal OPE in the UV at leading order, while the rhs is actually IR finite because of the mass gap in the closed-string sector of 't Hooft-planar PS. \par
Interestingly, the open/closed duality may be realized perturbatively in this model compatibly with the non-'t Hooft-planar perturbative renormalization, despite a nonperturbative mass gap occurs as opposed to the massless perturbation theory in Sec. \ref{1}, since the closed-string tree corresponding to the rhs of Eq. \ref{Th} would be UV divergent,
should PS plus SUSY matter in the fundamental representation exist as a canonical string theory. \par 
Yet, the resulting model has a positive non-'t Hooft planar beta function, and it flows to strong coupling in the UV, as opposed to a QCD-like theory.
Thus, the open/closed duality would be realized precisely because AF and the universally believed UV finiteness of the closed-string trees would be violated. \par
This is also the reason why the first no-go theorem (Subsect. \ref{E}) would not be contradicted, because its fundamental premises, AF and the universally believed UV finiteness of the closed-string trees would be violated. \par
Hence, in the large-$N$ 't Hooft limit of PS with SUSY matter in the fundamental representation, we have traded the potential perturbative consistency with the open/closed duality on the string side for a flow to strong coupling in the UV.\par
To avoid the UV flow to strong coupling -- at the price of increasing complications -- some PS variants \cite{PS} have been proposed, where $SU(N)$ $\mathcal{N}=1$ SUSY QCD is embedded into the IR of an UV finite $SU(N)$ $\mathcal{N}=2$ SUSY theory with $N_f$ SUSY massive matter fields in the fundamental representation by fine tuning $N_f$ with $N$ to get a vanishing beta function \cite{PS}.  \par
Because of the fine tuning, the finiteness of the $\mathcal{N}=2$ SUSY model extends to the finiteness of its large-$N$ limit in the Veneziano expansion (see Sec. 6 in \cite{MBR}) where the ratio $\frac{N_f}{N}$ is kept fixed compatibly with the fine tuning. Indeed, $\beta_0$ in the $SU(N)$ $\mathcal{N}=2$ SUSY model \footnote{In the $\mathcal{N}=2$ SUSY model only $\beta_0$ may be nonvanishing.} and the associated Veneziano-planar $\beta_0$, $\beta^{VP}_0$, both vanish (see Sec. 6 in \cite{MBR}), as opposed to the 't Hooft-planar $\beta_0$, $\beta^{P}_0$.\par
However, in \cite{PS} it has been claimed that for $N_f \sim N$ "there is as yet no dual string description", unsurprisingly because the Veneziano expansion \cite{V0} inevitably arises for large $N_f \sim N$, as opposed to the 't Hooft expansion.\par
The renormalization properties of the Veneziano expansion in QCD and $\mathcal{N}=1$ SUSY QCD have been worked out in \cite{MBR}. 
The compatibility of the Veneziano expansion in large-$N$ QCD with the open/closed duality is outside the scope of this paper, but it has been briefly discussed in \cite{HEP2017}.
Presently, no string model solving conjecturally for the Veneziano expansion has been shown to satisfy Eqs. \ref{11} and \ref{21}, which are implied by AF. \par
Finally, for a specific low value of $N_f$ there exists another PS variant \cite{PS} that is UV finite and whose large-$N$ 't Hooft-planar limit is UV finite as well, assumedly with a mass gap in the 't Hooft-planar closed-string sector. It is an $Sp(N)$ $\mathcal{N} = 2$ SUSY model \cite{A1,A2} deformed by mass terms \cite{PS}, with one multiplet in the antisymmetric tensor and four multiplets in the fundamental representation.\par
Then, the perturbative low-energy theorem (Sec. \ref{1}) demonstrates the perturbative compatibility with the open
/closed duality as above, because of the vanishing of the beta function both in the $Sp(N)$ $\mathcal{N} = 2$ SUSY model and in its large-$N$ 't Hooft-planar limit. In this case, the consistency of the open/closed duality with the mass gap occurs rather trivially, as no divergence arises both on the open and closed side of the duality, because of the IR and UV finiteness in Eqs. \ref{OPE}, \ref{Th}, \ref{div}.\par
By summarizing, none of the aforementioned examples satisfies the QCD-like properties 1, 2 and 3, and therefore none of them contradicts the no-go theorems. \par
In these examples the perturbative low-energy theorem is compatible at lower orders with the open/closed duality precisely because of the lack of AF. \par
We should also add that, even by limiting ourselves to closed strings only, for which no consistency issue with AF arises in the large-$N$ 't Hooft expansion (Subsec. \ref{D}), no canonical string model presently exists \cite{MBM,MBN} whose two-point correlator for $\Tr F^2$ has been shown to factorize over an infinite tower of massive states and to agree asymptotically in the UV with the result implied by AF in Eq. \ref{11} and \ref{21}, as it must occur in the 't Hooft-planar sector of a confining asymptotically free QCD-like theory \cite{MBM,MBN}. \par

\section{Acknowledgments}
We would like to thank Elisabetta Pallante and Gabriele Veneziano for the helpful comments.
We would like to thank Matteo Becchetti for checking the signs in Eq. \ref{OPE} and Alessandro Pilloni for working out the duality plot.

\end{document}